\pdfoutput=1
\documentclass[12pt,a4paper]{article}

\usepackage{ifthen} 
\newboolean{pdflatex}
\setboolean{pdflatex}{true} 

\newboolean{articletitles}
\setboolean{articletitles}{true} 

\newboolean{uprightparticles}
\setboolean{uprightparticles}{false} 

\def\paperauthors{LHCb collaboration} 
\def\paperasciititle{Technology developments for LHCb Upgrade II} 
\def\papertitle{Technology developments for\\ LHCb Upgrade II} 
\def\paperkeywords{ {LHCb}} 
\def\papercopyright{\the\year\ CERN for the benefit of the LHCb collaboration} 
\def\paperlicence{CC BY 4.0 licence}
\def\paperlicenceurl{https://creativecommons.org/licenses/by/4.0/}

\usepackage[top=1in, bottom=1.25in, left=1in, right=1in]{geometry}

\usepackage{microtype}
\usepackage{lineno}  
\usepackage{xspace} 
\usepackage{caption} 

\usepackage{graphicx}  
\usepackage{color}
\usepackage{colortbl}
\graphicspath{{./figs/}} 

\usepackage{amsmath} 
\usepackage{amssymb}
\usepackage{amsfonts}
\usepackage{upgreek} 

\usepackage{multirow}

\newcommand*\patchAmsMathEnvironmentForLineno[1]{%
\expandafter\let\csname old#1\expandafter\endcsname\csname #1\endcsname
\expandafter\let\csname oldend#1\expandafter\endcsname\csname
end#1\endcsname
 \renewenvironment{#1}%
   {\linenomath\csname old#1\endcsname}%
   {\csname oldend#1\endcsname\endlinenomath}%
}
\newcommand*\patchBothAmsMathEnvironmentsForLineno[1]{%
  \patchAmsMathEnvironmentForLineno{#1}%
  \patchAmsMathEnvironmentForLineno{#1*}%
}
\AtBeginDocument{%
\patchBothAmsMathEnvironmentsForLineno{equation}%
\patchBothAmsMathEnvironmentsForLineno{align}%
\patchBothAmsMathEnvironmentsForLineno{flalign}%
\patchBothAmsMathEnvironmentsForLineno{alignat}%
\patchBothAmsMathEnvironmentsForLineno{gather}%
\patchBothAmsMathEnvironmentsForLineno{multline}%
\patchBothAmsMathEnvironmentsForLineno{eqnarray}%
}

\usepackage[pdftex,
            pdfauthor={\paperauthors},
            pdftitle={\paperasciititle},
            pdfkeywords={\paperkeywords},
]{hyperref}

\usepackage{hyperxmp}

\hypersetup{pdfcopyright={Copyright (C) \papercopyright}}
\hypersetup{pdflicenseurl={\paperlicenceurl}}

\usepackage[bottom,flushmargin,hang,multiple]{footmisc}

\usepackage[all]{hypcap} 

\usepackage{xspace} 
\usepackage{upgreek}

\def\MagUp {\mbox{\em Mag\kern -0.05em Up}\xspace}

\ifthenelse{\boolean{uprightparticles}}%
{

 \def\PDelta      {\ensuremath{\Delta}\xspace}                 
 \def\PXi         {\ensuremath{\Xi}\xspace}                 
 \def\PLambda     {\ensuremath{\Lambda}\xspace}                 
 \def\PSigma      {\ensuremath{\Sigma}\xspace}                 
 \def\POmega      {\ensuremath{\Omega}\xspace}                 
 \def\PUpsilon    {\ensuremath{\Upsilon}\xspace}
 \let\oldPi\Pi
 \def\PPi         {\ensuremath{\oldPi}\xspace}

 \def\PB      {\ensuremath{\mathrm{B}}\xspace}                 
                  
 \def\PD      {\ensuremath{\mathrm{D}}\xspace}

 \def\PK      {\ensuremath{\mathrm{K}}\xspace}

 \def\Pi      {\ensuremath{\mathrm{i}}\xspace}

 \def\Ps      {\ensuremath{\mathrm{s}}\xspace}

 \def\thebaroffset{0.0em}
}
{

 \mathchardef\PDelta="7101
 \mathchardef\PXi="7104
 \mathchardef\PLambda="7103
 \mathchardef\PSigma="7106
 \mathchardef\POmega="710A
 \mathchardef\PUpsilon="7107
 \mathchardef\PPi="7105
                  
 \def\PB      {\ensuremath{B}\xspace}                 
                  
 \def\PD      {\ensuremath{D}\xspace}

 \def\PK      {\ensuremath{K}\xspace}

 \def\Pi      {\ensuremath{i}\xspace}

 \def\Ps      {\ensuremath{s}\xspace}

 \def\thebaroffset{0.18em}
}
\newcommand{\offsetoverline}[2][\thebaroffset]{\kern #1\overline{\kern -#1 #2}}%

\makeatletter
\ifcase \@ptsize \relax
  \newcommand{\miniscule}{\@setfontsize\miniscule{4}{5}}
\or
  \newcommand{\miniscule}{\@setfontsize\miniscule{5}{6}}
\or
  \newcommand{\miniscule}{\@setfontsize\miniscule{5}{6}}
\fi
\makeatother

\DeclareRobustCommand{\optbar}[1]{\shortstack{{\miniscule (\rule[.5ex]{1.25em}{.18mm})}
  \\ [-.7ex] $#1$}}

\def\squark    {{\ensuremath{\Ps}}\xspace}

\def\KorKbar {\kern \thebaroffset\optbar{\kern -\thebaroffset \PK}{}\xspace}

\def\D       {{\ensuremath{\PD}}\xspace}

\def\DorDbar {\kern \thebaroffset\optbar{\kern -\thebaroffset \PD}\xspace}

\def\Dp      {{\ensuremath{\D^+}}\xspace}
\def\Dm      {{\ensuremath{\D^-}}\xspace}

\def\DpDm    {\ensuremath{\Dp {\kern -0.16em \Dm}}\xspace}

\def\B       {{\ensuremath{\PB}}\xspace}

\def\BorBbar {\kern \thebaroffset\optbar{\kern -\thebaroffset \PB}\xspace}

\def\Bd      {{\ensuremath{\B^0}}\xspace}

\def\BdorBdbar {\kern \thebaroffset\optbar{\kern -\thebaroffset \Bd}\xspace}

\def\Bs      {{\ensuremath{\B^0_\squark}}\xspace}

\def\BsorBsbar {\kern \thebaroffset\optbar{\kern -\thebaroffset \Bs}\xspace}

\def\Y#1S{\ensuremath{\PUpsilon{(#1S)}}\xspace}

\def\LorLbar     {\kern \thebaroffset\optbar{\kern -\thebaroffset \PLambda}\xspace}

\def\AT#1     {\ensuremath{A_{\mathrm{T}}^{#1}}\xspace}           

\def\C#1      {\ensuremath{\mathcal{C}_{#1}}\xspace}                       
\def\Cp#1     {\ensuremath{\mathcal{C}_{#1}^{'}}\xspace}                    
\def\Ceff#1   {\ensuremath{\mathcal{C}_{#1}^{\mathrm{(eff)}}}\xspace}        
\def\Cpeff#1  {\ensuremath{\mathcal{C}_{#1}^{'\mathrm{(eff)}}}\xspace}       
\def\Ope#1    {\ensuremath{\mathcal{O}_{#1}}\xspace}                       
\def\Opep#1   {\ensuremath{\mathcal{O}_{#1}^{'}}\xspace}                    

\newcommand{\nospaceunit}[1]{\ensuremath{\text{#1}}}       
\newcommand{\aunit}[1]{\ensuremath{\text{\,#1}}}       

\newcommand{\tev}{\aunit{Te\kern -0.1em V}\xspace}
\newcommand{\gev}{\aunit{Ge\kern -0.1em V}\xspace}
\newcommand{\mev}{\aunit{Me\kern -0.1em V}\xspace}
\newcommand{\kev}{\aunit{ke\kern -0.1em V}\xspace}
\newcommand{\ev}{\aunit{e\kern -0.1em V}\xspace}
 
\newcommand{\mevc}{\ensuremath{\aunit{Me\kern -0.1em V\!/}c}\xspace}
\newcommand{\gevc}{\ensuremath{\aunit{Ge\kern -0.1em V\!/}c}\xspace}
\newcommand{\mevcc}{\ensuremath{\aunit{Me\kern -0.1em V\!/}c^2}\xspace}
\newcommand{\gevcc}{\ensuremath{\aunit{Ge\kern -0.1em V\!/}c^2}\xspace}

\def\m    {\aunit{m}\xspace}

\def\cm   {\aunit{cm}\xspace}

\def\mm   {\aunit{mm}\xspace}

\def\mum  {\ensuremath{\,\upmu\nospaceunit{m}}\xspace}

\def\nm   {\aunit{nm}\xspace}

\def\sec  {\ensuremath{\aunit{s}}\xspace}

\def\ps   {\ensuremath{\aunit{ps}}\xspace}

\def\mhz  {\ensuremath{\aunit{MHz}}\xspace}

\def\MRad {\aunit{MRad}\xspace}

\def\gsim{{~\raise.15em\hbox{$>$}\kern-.85em
          \lower.35em\hbox{$\sim$}~}\xspace}
\def\lsim{{~\raise.15em\hbox{$<$}\kern-.85em
          \lower.35em\hbox{$\sim$}~}\xspace}

\def\tell1  {TELL1\xspace}
\def\ukl1   {UKL1\xspace}

\newcommand{\etc}{\mbox{\itshape etc.}\xspace}

\newcommand{\lhcborcid}[1]{\href{https://orcid.org/#1}{\hspace*{0.1em}\raisebox{-0.45ex}{\includegraphics[width=1em]{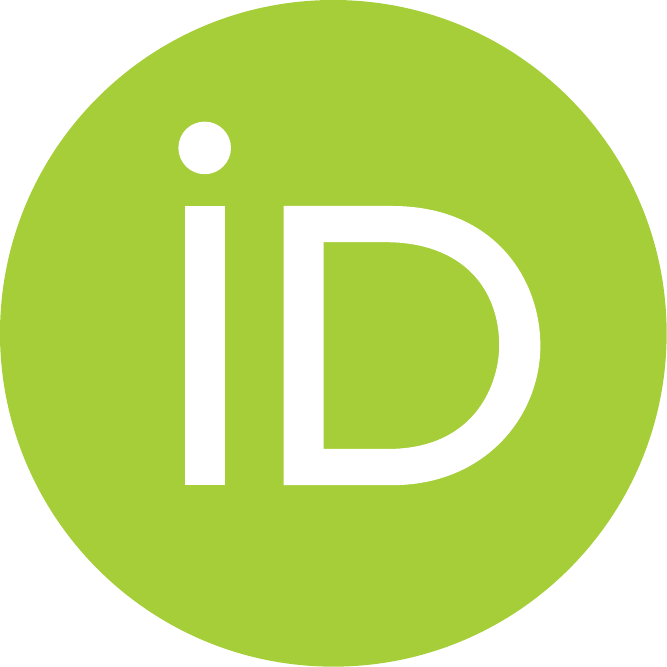}}}}

\usepackage{cite} 
\usepackage{mciteplus}

\usepackage{pdflscape}
\usepackage{etoc}

\usepackage{comment}
\usepackage{tikz}
\usetikzlibrary{shapes}
\usepackage{pgfgantt}
\usepackage{pdflscape}

\begin{document}

\renewcommand{\thefootnote}{\fnsymbol{footnote}}
\setcounter{footnote}{1}

\begin{titlepage}
\pagenumbering{roman}

\vspace*{-1.5cm}
\centerline{\large EUROPEAN ORGANIZATION FOR NUCLEAR RESEARCH (CERN)}
\vspace*{1.5cm}
\noindent
\begin{tabular*}{\linewidth}{lc@{\extracolsep{\fill}}r@{\extracolsep{0pt}}}
\ifthenelse{\boolean{pdflatex}}
{\vspace*{-1.5cm}\mbox{\!\!\!\includegraphics[width=.14\textwidth]{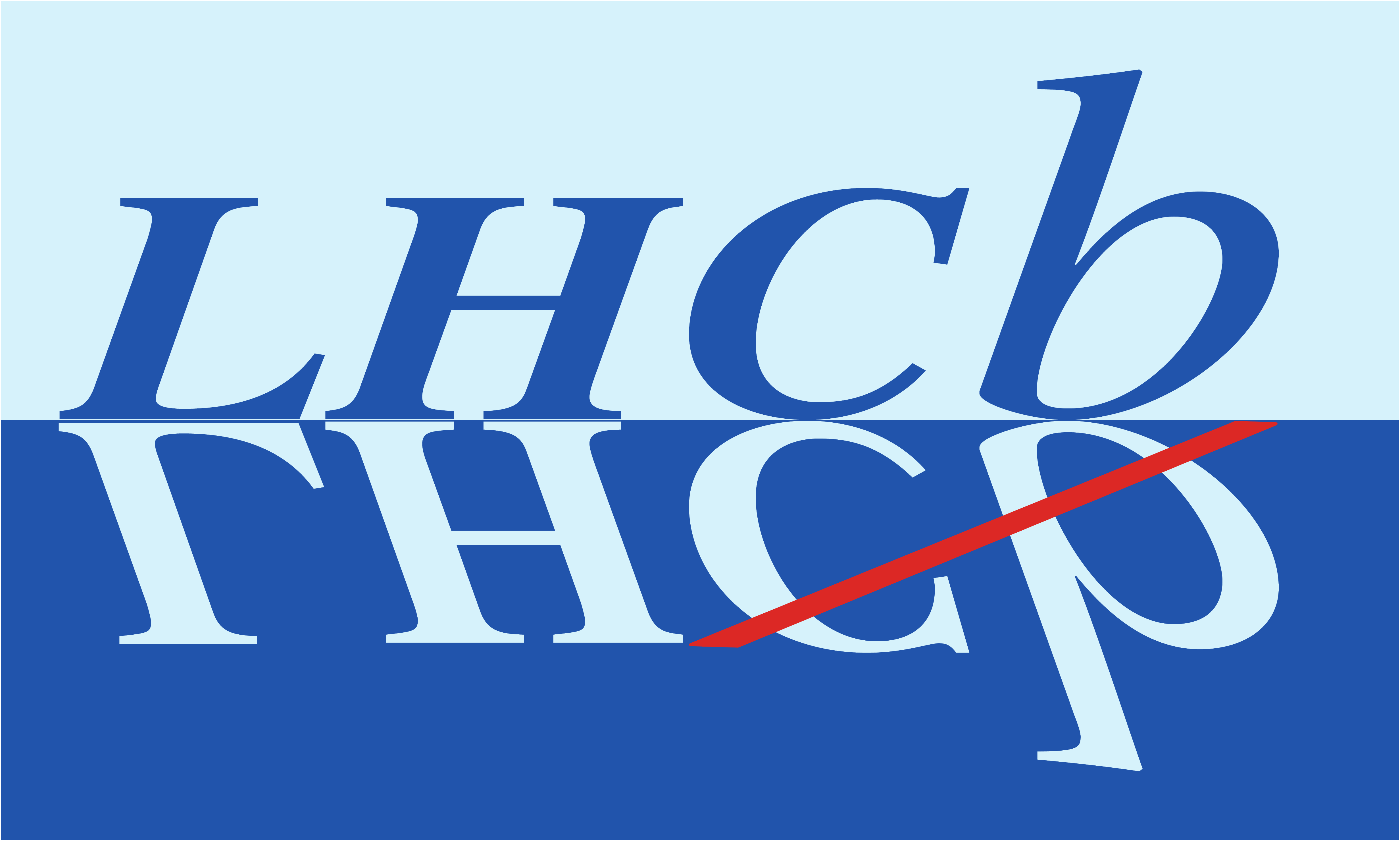}} & &}%
{\vspace*{-1.2cm}\mbox{\!\!\!\includegraphics[width=.12\textwidth]{figs/lhcb-logo.eps}} & &}%
\\
 & & LHCb-PUB-2025-002 \\  
 & & March 31, 2025 \\ 
 & & \\
\end{tabular*}

\vspace*{2.0cm}

{\normalfont\bfseries\boldmath\huge 
\begin{center}
    \papertitle \\
    \vspace*{0.5cm}
  {\normalsize Input to the European Particle Physics Strategy Update 2024--26}
\end{center}
}

\vspace*{1.0cm}

\begin{center}
\paperauthors\footnote{
    Contact authors: 
    Vincenzo Vagnoni (\href{mailto:vincenzo.vagnoni@cern.ch}{vincenzo.vagnoni@cern.ch}),
    Tim Gershon (\href{mailto:tim.gershon@cern.ch}{tim.gershon@cern.ch}),
    Giovanni Punzi (\href{mailto:giovanni.punzi@cern.ch}{giovanni.punzi@cern.ch})
}
\end{center}

\vspace{\fill}

\begin{abstract}
    \noindent
    A major LHCb detector upgrade will be installed during long shutdown~4 (LS4) of the CERN Large Hadron Collider.
    The experiment will operate at a maximum luminosity of up to $1.5\times 10^{34}\cm^{-2}\sec^{-1}$, with acceptance covering a pseudorapidity range close to the beamline.
    The detector will therefore experience extremely high particle fluences.
    In order to carry out the LHCb physics programme, technologies are being developed that can withstand the high rates and associated radiation damage while also providing excellent resolution in both space and time.
    The amount of data to be processed in the online computing and trigger system is also unprecedented.  
    In this document, the technology developments that are necessary to realise this programme are summarised.
\end{abstract}

\vspace{\fill}

{\footnotesize 
\centerline{\copyright~\papercopyright. \href{\paperlicenceurl}{\paperlicence}.}}
\vspace*{2mm}

\end{titlepage}



\renewcommand{\thefootnote}{\arabic{footnote}}
\setcounter{footnote}{0}

\cleardoublepage


\pagestyle{plain} 
\setcounter{page}{1}
\pagenumbering{arabic}



\section{Introduction}
The successful operation and rich physics harvest of LHCb during Run~1 and Run~2 of the LHC vindicated the concept and design of a dedicated heavy-flavour physics experiment at a hadron collider, showing potential going well beyond the low-luminosity regime. 
This has inspired the collaboration to upgrade the experiment's capabilities, to allow full exploitation of the unique physics opportunities offered by the unprecedented rate of heavy quarks that LHC collisions produce at full luminosity.
While the initial experiment operated at an instantaneous luminosity of $2\times 10^{32}\cm^{-2}\sec^{-1}$, the Upgrade~I of the detector has recently been proven to work at its design luminosity of $2\times 10^{33}\cm^{-2}\sec^{-1}$.
The LHCb collaboration is now designing for another order of magnitude increase, aiming at luminosities in excess of $10^{34}\cm^{-2}\sec^{-1}$, with its Upgrade~II project. 

\begin{figure}[!b]
    \centering
    \begin{minipage}{0.45\textwidth} 
        \centering
        \includegraphics[width=\linewidth]{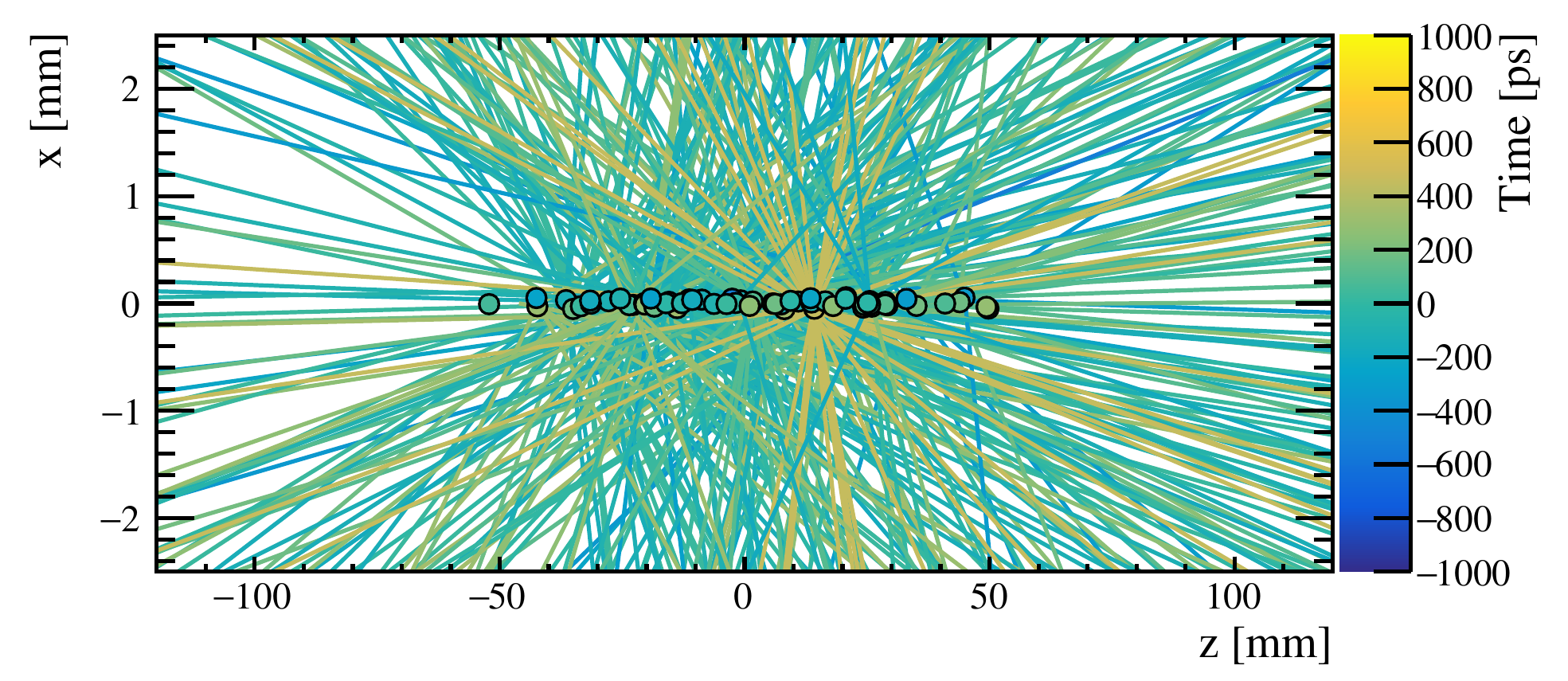}
        \vspace{0.01cm} 
        \includegraphics[width=\linewidth]{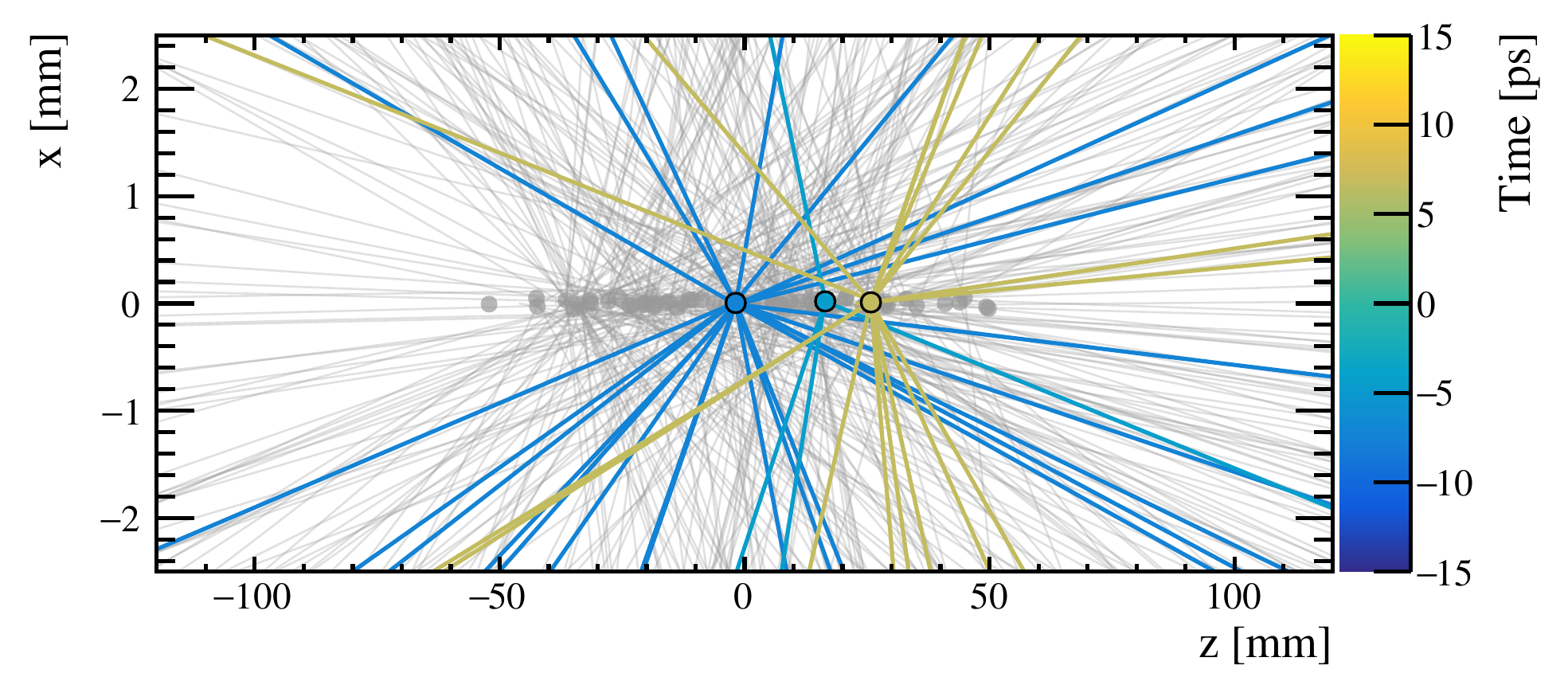}
    \end{minipage}
    \hfill
    \begin{minipage}{0.52\textwidth} 
        \centering
        \includegraphics[width=\linewidth]{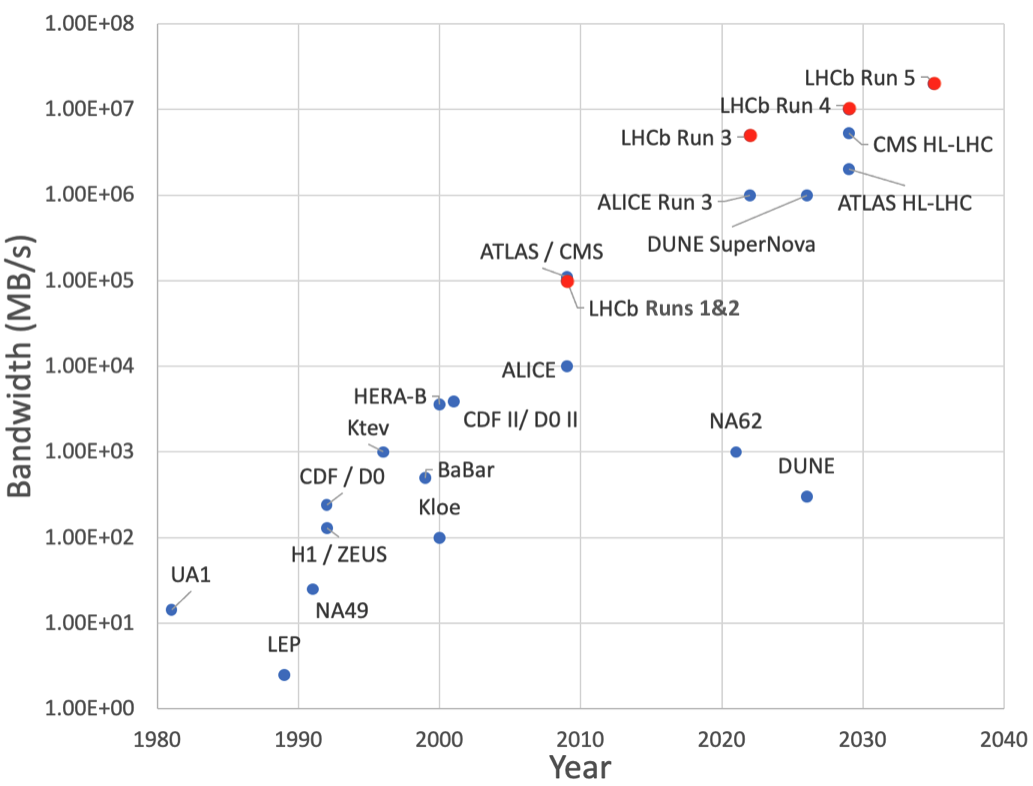}
    \end{minipage}
    
    \caption{(Left) Tracks in a typical bunch-crossing at LHCb in Upgrade-II (42 vertices), coloured according to time of production. The bottom plot highlights a specific 30\ps time window. 
    (Right) Evolution of experiment data rates in High Energy Physics as a function of time [graph courtesy of A.~Cerri, U.~of Sussex].}
    \label{fig:data-rates}
\end{figure}

Precision flavour physics requires accurate determinations of the positions and momenta of particles, as well as their identification.
Providing all these capabilities down to the low transverse momenta typical of the bulk of charm and beauty production, within the dense environment of hadron collisions, is a special challenge for the Upgrade~II project, imposing strong requirements beyond the needs of general-purpose experiments running at the same luminosity. 
They range from extremely good pattern recognition in a crowded environment, demanding very high detector granularity in both space and time, to extremely large amounts of sophisticated and custom-designed data processing~(Figure~\ref{fig:data-rates}).
All of this needs to be achieved within ever-increasing constraints of environmental viability. 
The present document is intended to provide an executive summary of the main lines of technological development ongoing within the context of LHCb Upgrade~II. It is organized in three broad areas: tracking, particle identification, and data processing. 
Full details about the Upgrade~II project can be found in the Framework TDR~\cite{LHCb-TDR-023}, and the Scoping Document~\cite{LHCb-TDR-026}. Additional details on data processing can be found in a separate submission to the present European Particle Physics Strategy Update~\cite{LHCb-PUB-2025-004}.

\section{Tracking technology}

Outstanding tracking performance will be critical to the success of LHCb Upgrade~II. The system must be capable of triggering and reconstructing primary, secondary and tertiary vertices of beauty and charm hadrons and delivering excellent momentum resolution across the full angular acceptance and momentum range, in very dense conditions~(Figure~\ref{fig:data-rates} left). 

The tracking system will look superficially similar to that of LHCb Upgrade~I, with a silicon pixel detector around the interaction region and large-area tracking detectors in front of and behind the magnet, with the addition of tracking stations inside the magnet to extend acceptance for low-momentum tracks.   However, in order to be able to deliver an acceptable signal-to-noise ratio and the efficiency required for track reconstruction in the trigger at the vastly increased occupancies at high luminosity, the tracking performance at Upgrade~II must in general match or surpass that of Upgrade~I.  The key enhancements that will make this possible are the addition of timing measurements at the order of tens of picoseconds at the level of the vertex detector, and the implementation of large-area pixel detectors in the downstream tracking, to complement the enhanced scintillating fibre detector.  Both represent an important stepping stone for the HEP community, in pushing forward technologies which are at the forefront of interest for the longer term future. 

Coming first to the hybrid pixel vertex detector, the operation close to the beamline at such high luminosities will lead to data rates of hundreds of kHz per pixel, with the corresponding need for huge ASIC data bandwidths, and the requirement to withstand fluences of up to $6 \times 10^{16}\,1\mev\,n_{\rm eq}/\!\cm^2$.  This must be achieved with a pixel-hit resolution which is better than that of the current VELO Upgrade~I detector, very high efficiency (to avoid adding layers) and a reduced material budget, in spite of the power and cooling challenges posed by the functionality being demanded of the ASIC and sensor.  Simulation studies have shown that the leap in performance needed for Upgrade~II will only be possible by the addition of per-pixel timing measurements.  The target is to reach less than 50\ps per measurement (30\ps contributions from the sensor and ASIC individually), leading to a 20\ps track time-stamping capability.  

Developing the technology to achieve this is well aligned with the global goals of the HEP community.  Timing with tracking was pioneered by the NA62 experiment\cite{Rinella_2019}, and has been carried forward by ATLAS and CMS which plan time-stamping tracking layers for the HL-LHC upgrades~\cite{atlas_timing,cms_timing}.  LHCb is an opportunity for a first step towards incorporating timing information for a full scale system in the real-time event reconstruction, paving the way for full 4D tracking detectors in the ensuing decades, which will be mandatory, for instance, for FCC-hh detectors.

On the sensor side, achieving the required temporal and spatial resolution at full efficiency for all track angles and over the full detector lifetime will be a crucial technological development.  Sensor timing technologies include thin planar sensors, Low Gain Avalanche Diodes, or possibly alternative materials; however the most promising technology
is given by 3D silicon sensors, which has excellent radiation hardness potential and for which the use (without timing) has been successfully consolidated in LHC experiments. In order to deliver the required performance, the 3D electrode design must be improved to deliver the most uniform timing performance at small pixel pitch as well as delivering sufficient signal to keep the electronics timing jitter under control.  The R\&D is currently focused on the manufacturing techniques for such sensors.

On the ASIC side, the LHCb Upgrade~II developments will use previous achievements as a springboard, for example the 65\nm ASIC developments for the ATLAS/CMS timing layers or the CLICpix small pixels ASICs~\cite{clicpix}. The main functionalities that should be added relate to the very high data rate and data transmission considerations, the power distribution design which is essential for achieving a uniform timing resolution over a full scale chip, the need for a certain amount of pixel processing on chip, and radiation hardness.  It will be mandatory to move to the 28\nm technology node, where experience is currently being gained in HEP, and for which the ASIC under study will be one of the first full scale demonstrators.  More complex technological options that could be considered include the use of through-silicon vias and vertical integration, such as 3D stacking.  A particularly exciting development on the horizon is silicon photonics~\cite{troska_pho}, which offers an ultra-low-mass solution highly integrated with the front-end electronics and capable of handling very high data rates. Intensive R\&D is ongoing on potential ASIC designs, and whatever mix of technologies is employed will be an important demonstrator for future projects, while being strongly aligned with industry trends.

The Upgrade~II vertex detector will be installed as a moveable system within the primary vacuum of the LHC.  For this reason it will also rely on cutting-edge technologies for vacuum-compatible materials such as carbon fibre elements, surface coatings that minimise secondary electron yields, and solutions for movements or possible detector replacements in high radiation environments.  These are also topics of general interest for HEP.

Another key technological advancement will be the extension of the use of pixels to larger detectors, covering $\mathcal{O}(10\m^2)$. 
This will enable standalone track reconstruction in the Upstream tracker (UT), located in the fringe field before the magnet, allowing 
elimination of ghosts even at very high occupancy.  The downstream tracker, dubbed Mighty Tracker (MT),  currently based on scintillating fibre technology, will move to a mixed solution, with the inner part equipped with pixels, allowing to cope with the high particle flux and integrated radiation dose. 
Both systems will need bunch-crossing tagging, hence a time resolution of a few nanoseconds, and to sustain radiation doses of up to $3 \times 10^{15}\,1\mev\,n_{\rm eq}/\!\cm^2$ (UT) and $6 \times 10^{14}\,1\mev\,n_{\rm eq}/\!\cm^2$ (MT). All this is required while keeping each layer highly efficient and contributing less than $1\%$ of a radiation length.  That poses a particular challenge for the MT, for which the mechanics and services have to be brought in to the centre of the scintillating fibre detector.  

The development efforts necessary to deliver these large pixel detectors are currently focussed on Monolithic Active Pixel Sensors (MAPS). This technology, which combines sensor and readout electronics on the same piece of silicon, provides a low-power and low-mass solution.  The availability of commercial CMOS processes with quadruple-well technologies to allow for full CMOS circuitry inside the pixel cell, as well as high resistivity substrate wafers to enable depleting either a part or the full sensing volume, have accelerated the progress in sensor development. MAPS are of great interest for HEP and have recently reached the technical maturity needed for large scale systems.  An early pioneer was the STAR experiment at RHIC~\cite{star_rhic}; the Inner Tracker Upgrade of ALICE is currently running at the LHC~\cite{alice_maps}; a future ALICE barrel tracker is planned with curved and stitched sensors~\cite{alice_run4}; beyond this, projected MAPS use cases include EIC detectors, DECAL for the ILC~\cite{ilc_decal} and the FCC-ee IDEA detector~\cite{tracking_fcc}. This places the LHCb upgrade firmly in line with the trends in HEP, and taking a major step for the community by constructing the first radiation-hard MAPS system at the LHC.   

The R\&D for LHCb MAPS is focused on achieving substantial depletion in the sensing volume, improving time resolution and radiation resistance without increasing power consumption. The developments for the required performance currently focus on the high-voltage (HV-CMOS) technology. The HV-CMOS approach builds on developments for Mu3e experiment, a possible ATLAS pixel upgrade, and on collaborations with the DRD3 community. These 180\nm and 150\nm CMOS processes demonstrate good intrinsic radiation hardness, and a pixel size of order $50 \times 150\mum^2$ is possible.   

For the scintillating fibre part of the MT tracker, 
new scintillators based on nano-organic luminescence (NOL) materials, which are green-emitting and may suffer less transparency loss than the current blue-emitting fibres, are being studied. Also, the Silicon PhotoMultipliers (SiPMs) used for fibre readout will need to be operated at cryogenic temperatures to ensure a dark rate count compatible with high luminosity operation, and allow lower thresholds to enhance efficiency. This will also be helped by the use of micro-lenses, concentrating the light from the fibres onto the active region of the SiPM. 

Finally, a new tracking component --- the Magnet Stations (MS), consisting of panels of triangular scintillating bars with a total coverage of about $14\m^2$ --- will be placed inside the magnet to extend LHCb's low-momentum acceptance.    
The scintillating light is collected by 1\mm wavelength shifters embedded inside the bars and guided outside the magnet with clear fibres, all outside the LHCb acceptance.  This subdetector poses interesting technological challenges for cooling, radiation hardness and readout, and for implementation of clustering and pattern recognition on next-generation FPGAs.

\section{Particle-identification technologies}
The RICH and TORCH detectors are key components of the particle identification (PID) system for LHCb Upgrade~II, designed to provide precise discrimination between different charged hadrons ($\pi/K/p$) across a wide momentum range. 
These detectors must operate reliably under the extreme conditions of future LHC runs, including high photon fluxes and significant radiation exposure, and ensure long-term stability and robust performance.
For the RICH detectors, achieving and maintaining excellent PID performance requires advancements in photon detection efficiency, substantial improvements in single-photon Cherenkov angle precision, better granularity and the use of timing measurements for photons, to reject backgrounds and to perform accurate photon association to particle tracks in spite of the crowded environment. The TORCH detector depends on precise timing measurements for effective particle identification in the low-momentum range.
To meet these challenges, several cutting-edge technological advancements are needed.  
This is complemented by an effort on system sustainability and strict control of systematic uncertainties, that might otherwise become relevant at high luminosity.

The requirements for photon detectors are demanding. They must withstand photon counting rates of the order of several MHz/mm$^2$, be resilient to levels of radiation up to $10^{14}\,n_{\rm eq}/\!\cm^{2}$, and withstand a total ionizing dose up to a few $\!\MRad$. In addition, they must provide high photon detection efficiency to maximize photon yield, allow covering areas of several m$^2$, achieve single-photon timing resolutions of ${\cal O}(10)\ps$, and feature sub-millimeter pixel sizes for high spatial resolution with large filling factor. R\&D efforts are focused on advancing photodetector technologies tailored to the specific needs of the RICH and TORCH systems. For the RICH detector, the emphasis is on SiPMs and Large Area Picosecond Photon Detectors (LAPPDs), while the TORCH detector's baseline technology is pixellated Microchannel Plate Photomultiplier Tubes (MCP-PMTs). 

SiPMs combine compact design, low operational voltage, robustness to magnetic fields, and relatively low cost. Their high photon detection efficiency (PDE) makes them ideal for single-photon detection, maximizing photon yield and enabling precise Cherenkov angle reconstruction. Despite their many advantages, SiPMs face issues that ongoing R\&D aims to address. A major one is their sensitivity to radiation-induced damage, resulting in high dark count rates and reduced single-photon resolution. Efforts to overcome these challenges include developing new SiPMs with increased radiation hardness and implementing mitigation strategies such as radiation shielding, annealing, and operation at low temperatures. The development of effective radiation shielding could prevent damage in experiments where SiPMs can be positioned outside the acceptance region. In-situ annealing is an effective method for mitigating radiation damage but requires designs and materials capable of withstanding temperature excursions, including the photodetector housing structure, which must ensure adequate thermal separation from the surrounding environment. Advanced cooling systems are needed to reduce the dark count rate sensitivity to irradiation, ensuring long-term efficient single-photon detection. Significant benefits could be achieved by developing or adapting localized cooling technologies specifically designed for solid-state photosensors.  Additionally, while SiPMs already exhibit excellent timing characteristics, ongoing R\&D aims to further improve timing resolution, a crucial requirement for applications such as the TORCH detector. Furthermore, in high-occupancy environments, the development of fast-recovery devices is critical.  
MCP-PMTs provide excellent radiation tolerance and timing resolution (tens of picoseconds). However, both their rate capability and integrated charge capability must be improved by more than an order of magnitude compared to the current state-of-the-art, along with enhanced PDE.
For the RICH detectors, a possible alternative to SiPM technology is given by LAPPDs. These devices offer large detection areas and similarly excellent timing resolution. However, LAPPDs face significant challenges, similar to those for MCP-PMTs, that must be addressed to make them suitable for high-luminosity operation.

The development of front-end electronics capable of precise time-stamping of Cherenkov photon signals, with a resolution in the tens of \!\ps, is crucial for both the RICH and TORCH detectors, for improving overall performance under high-luminosity conditions and for accurate measurement of the time of flight of particles by the TORCH system.
A front-end ASIC is under development that combines analog and digital functionalities into a single compact, radiation-hard, and low-power multi-channel chip (``FastRICH''). This design allows for a wide input signal dynamic range for coupling to different types of photon sensors and incorporates a constant-fraction discriminator (CFD) and a time-to-digital converter (TDC), enabling the needed timing precision and providing data-compressed output to minimize bandwidth. Further R\&D will evaluate compatibility with photon sensors such as SiPMs operated across a wide temperature range, develop techniques for managing the thermal gradients produced by heat dissipated into small volumes, and integrating mechanical, thermal, electrical and optical functions at high granularities.

Intense efforts are underway to replace the fluorocarbon gases currently used as Cherenkov radiators in RICH detectors with more sustainable alternatives having lower Global Warming Potential (GWP), while still matching their refractive index, photon yield and chromatic dispersion.  
The importance of this transition is heightened by upcoming regulations that are expected to further restrict the use of fluorocarbons, and thus are likely to prevent their use in future experiments. Technological advancements are essential to enable greater flexibility in selecting the refractive index by blending different gases and potentially operating with variable compositions to mitigate systematic effects. This requires improved methods for real-time monitoring of gas radiator properties, whether these be an evolution of the present techniques that exploit specifically selected beam events, and/or novel, fully detector-independent measurements of refractive index, possibly including measurement of inhomogeneities within the volume. 
In addition, the use of highly polished quartz radiator bars is critical for the TORCH detector. These radiators, spanning several tens of square meters in area, must meet stringent specifications, including (sub)-nanometer surface roughness and a high degree of planarity, to ensure optimal Cherenkov photon generation and minimal scattering.  
Achieving these standards over large areas necessitates advancements in quartz fabrication technologies and the development of cost-effective manufacturing processes. Furthermore, precise characterization techniques are also needed to verify the quality of the optical components, including planarity, optical transmission, scattering, and resistance to radiation darkening. In addition, there is need for quartz windows to separate the cooled photodetector volumes from the gas radiators in RICH detectors, which typically condense at temperatures a few tens of degrees Celsius below zero, and preserve their homogeneity. Active thermal control for these windows is being studied to address this issue effectively.

New optical geometries for improved RICH performance are also being explored through simulations. These require lightweight mirror technology, utilizing cost-effective materials. 
Precision and lightweight mechanical supports are also essential for mechanical integration with the global detector system. 
These advancements also benefit the development of full-coverage detectors at future colliders such as the FCC. 
Finally, improved Cherenkov angle accuracy necessitates an enhanced understanding and control of systematic effects. An advanced, comprehensive calibration, monitoring, and alignment system will be developed to maintain consistent performance and control systematic uncertainties under varying experimental conditions of the ${\cal O}(10^6)$ channels involved.

In addition to the RICH and TORCH detectors, the particle identification capabilities of the LHCb Upgrade~II detector rely on its sampling electromagnetic calorimeter (dubbed PicoCal) and its muon system.
For the latter, the main challenges are the particle flux rate, reaching up to  
$\mathcal{O}(1)\mhz/\!\cm^{2}$ in the inner regions, and the charge integrated over the lifetime of the upgraded experiment (up to $\sim1\,{\rm C}/\!\cm^{2}$).
To cope with the high rate, the inner regions of the muon stations are planned to be instrumented with 
\textmu{RWELL} detectors, a novel variant of micropattern gas detectors consisting of a copper-clad polyimide foil with a matrix of holes (wells) providing high-field regions for electron amplification.
To suppress discharges (thus enhancing the rate capability of the detector), the amplification stage is separated from the readout PCB by a thin Diamond-Like-Carbon (DLC) resistive film.
Further optimisation (tuning of the resistivity, grounding scheme, improvement of the uniformity, stable operation at high rate \etc) of the \textmu{RWELL} technology is an active area of R\&D.
The standard gas mixture currently used for muon detectors, \textmu{RWELL} and MWPC, includes  carbon tetrafluoride (CF$_4$) which has a high GWP.
The search for and validation of an environmentally friendly alternative able to guarantee equivalent performance represents an important target for environmental sustainability, not only for LHCb but also for the whole gas-based detector community. In case an adequate alternative will not be found, the greatest effort will be put in developing a highly efficient close-mode system to minimise CF$_4$ losses into the atmosphere.
In order to keep the readout rate at a manageable level, the readout scheme of the muon system needs a major redesign. 
This implies a significant increase (by a factor $\sim6$) in the number of readout channels with respect to the Upgrade~I detector, requiring the development of new front-end ASICs, and FPGA-based stub reconstruction in the back-end.

Timing capabilities of order $20\ps$ precision are crucial for pile-up mitigation in the PicoCal. In parallel, the current energy resolution of $\sigma(E)/E \approx 10\%/\sqrt{E} \oplus 1\%$ needs to be preserved.
The region of the PicoCal close to the beam axis will consist of Spaghetti Calorimeter (SpaCal) modules with scintillating fibres inserted in a dense absorber material. 
The innermost modules will have to sustain radiation doses of up to several $10^{15}\,1\mev\,n_{\rm eq}/\!\cm^{2}$. 
These will be made from tungsten absorbers with radiation-hard scintillating crystal fibres. In an ongoing R\&D campaign, the production of GAGG (Gadolinium Aluminium Gallium Garnet) fibres with decay times below 10 ns is being optimised. This technology would also be an interesting option for electromagnetic calorimetry at FCC-hh. As an alternative crystal material, YAG (Yttrium Aluminium Garnet) is being investigated.
The surrounding region will have modules with lead absorbers and radiation-tolerant organic scintillators. The PicoCal requirements are one of the reasons why interest in novel organic scintillating materials is growing recently. This includes studies of new fast green or orange emitters, and radiation-tolerant hosts beyond polystyrene. The corresponding R\&D provides synergies with developments aimed at Higgs factories.

Different technologies for the production of the SpaCal absorbers are being explored already in view of the ECAL enhancement during LS3 and will continue to be an important subject of R\&D. Use of 3D printing with tungsten is being pioneered in high energy physics for the innermost regions. A novel low-pressure casting technique for the lead absorbers is being optimised. An alternative solution based on a deep drilling approach is also being studied.
For the other regions of the PicoCal, further from the beam axis, the Shashlik technology used in the current LHCb electromagnetic calorimeter is being refined, in particular to improve the time resolution. This includes the replacement of the WLS fibres with faster materials.

Achieving the required time resolution will require careful optimisation not only of the absorber material and scintillators but of all elements in the detection and readout chain (light guides, PMTs, readout ASICs).
Hollow light guides made of 3D-printed plastic structures covered internally with reflective foils are being optimised and provide intrinsic radiation hardness. Photon detector candidates are being studied in test benches and on prototypes in test beam measurements. The development of two new front-end ASICs is ongoing in 65 nm technology, due to its long expected lifetime. The ICECAL65 targets energy measurements over an enlarged dynamic range. On the other hand, the SPIDER aims to provide the time measurements using waveform sampling in analogue memories.

\section{Data processing technology}
High Energy Physics experiments operating at the LHC are at the forefront of real-time computing today, and will remain so over the next 
decades during the machine's high-luminosity phase. The maximal scientific exploitation of the HL-LHC collider drives all the experiments to shift some or all real-time filtering tasks from fixed-function low-level trigger systems to flexible computing infrastructure enabling
physics-analysis-quality reconstruction and calibration in real or quasi-real time.

Already today the LHCb experiment is 
processing over two Terabytes of data per second in a heterogeneous data centre consisting of GPU and CPU processors, rates which will be reached by 
the ATLAS and CMS detectors in the HL-LHC phase. The second upgrade of LHCb, proposed for the mid-2030s, will push these data rates an order of magnitude 
higher, making it the biggest data processing challenge ever faced by High Energy Physics~(Figure~\ref{fig:data-rates}).

Performing a cost-effective, energy-efficient, interpretable, and maintainable real-time detector reconstruction at these data rates is one of 
the principal technological challenges facing the field over the next decade. As more and more physics-analysis-quality work is done near the
detector, the reusability of the real-time code and its interoperability with offline workflows become critical design criteria, on a par with more 
traditional considerations like cost-effectiveness and physics efficiency. For all these reasons, the data processing technologies and methods developed by the LHCb collaboration over the next decade will unquestionably act as a technology driver and catalyst for all future generations of HEP experiments.

In contrast to detector development, the data processing hardware used by HEP experiments mostly consists of off-the-shelf components. For the last two decades
this has been hugely beneficial, as the field benefitted from Moore's law and the ubiquitous nature of the x86 processing architecture for most high-level
tasks while developing its own custom ASIC- and FPGA-based solutions for near-detector processing. Today, however, the data processing hardware-landscape
is more complex and fragmented, with GPUs increasingly at the heart of commercial data processing, x86 facing increasing competition from more energy-efficient architectures like ARM and RISC-V, FPGAs being deployed as co-processors alongside both CPUs and GPUs in data centres, and even more specialised architectures appearing on the horizon. 
LHCb needs to exploit in the best possible way the most suitable technologies that will become available. This requires adapting and creatively reformulating its computational problems to work well with these new technologies. It must also take into account that sustainable computing policies are likely to become a key-element influencing the design of any new data centre. 
The optimal exploitation of emerging and future architectures will imply not only technological developments but also the ability to recruit, develop, and retain heterogeneous teams of physicists, software, firmware, and electronics engineers with the right mixture of skill-sets to address every facet of the task ahead of us. In fact, the building of these teams can be regarded as a by-product of this project that will outlive it and be an asset for the longer term future. 

All major HEP collaborations recognise the challenge of heterogeneous computing, but addressing it has required transforming legacy software stacks 
consisting of millions of lines of code, which are still mainly optimised for x86 architectures. This landscape is further complicated by the increasing 
proliferation of performance-oriented programming languages outside HEP, some of which like Rust and Julia are beginning to gain a foothold in our 
communities even if not yet in today's production workflows. 
Compounding this challenge is the fact that, while Moore's law is still
valid (transistor density on integrated circuits grows by about 30\% per 
year), the corresponding growth in compute power is only available in 
highly specialised forms. For example, the recent AI boom led to the
development of high-performance circuits dedicated to half-precision
floating-point operations. This makes the processing efficiency dependent
on the ability to exploit this specialised hardware, making generally
optimal solutions hard to design. Furthermore, the experience
gained during the design of LHCb's first upgrade demonstrated that the
computational cost of moving data into a specialised processing unit's memory can be comparable with the cost of processing it, which further complicates the design of data centres which can be efficient for a wide range of applications.

While formidable, these challenges are however also an opportunity for HEP. The scientific and technical content of LHCb problems -- accurate, reproducible, and interpretable exascale multidimensional inference -- is the same faced by industry and other scientific domains. Having to tackle these problems at an 
unprecedented scale has a strong potential to lead to solutions with an impact on the wider world.

The LHCb collaboration has been at the forefront of HEP data processing over the past 15 years. LHCb's data acquisition systems have always co-designed the computing, networking, storage, and software to be cost-effective and best suit
the computational problem at hand, and have promoted the use of commercial components 
as early as possible in the data acquisition chain. The collaboration pioneered the widespread use of quantised
multivariate algorithms~\cite{Gligorov:2012qt} from the beginning of Run~1; demonstrated the ability to perform the full offline-quality detector reconstruction, as well 
as detector alignment and calibration, in quasi-real time at the beginning of Run~2; and deployed a fully GPU-based first-level tracking trigger reconstructing soft-hadrons and electrons across the whole detector kinematic acceptance at the full LHC collision rate from the beginning of Run~3. 
Between Runs~2~and~3 LHCb also leveraged parallel-processing methodologies to improve the energy efficiency of its real-time processing by more than one order of magnitude~\cite{Aaij:2021oqw}. LHCb has also shown its ability to leverage these technological advances for physics analysis, in particular by pioneering the use of real-time reconstructed information in physics analyses during Run~2, together with the use of reduced data formats and automatic analysis tools. Thanks to these advances LHCb has been able to present~\cite{LHCb-TDR-026} a viable architecture for the second upgrade based on transferring the full detector 
reconstruction and data processing onto general-purpose GPU processors from Run~5 onwards. 

The collaboration remains engaged in a broad 
R\&D programme including novel processing architectures and data-centre scale as well as within-server interconnection technologies. These will provide safety in the case of significant shifts in the commercial markets, and position LHCb to take full advantage of new opportunities. In particular, LHCb will build and deploy~\cite{LHCb-TDR-025} the RETINA
system for performing early track reconstruction on FPGA boards in Run~4, which will allow it to gain valuable experience with this technology and 
facilitate the choice of optimal processing hardware for Run~5. In addition, the collaboration continues to expand its Allen~\cite{Aaij:2019zbu} framework for heterogeneous
processing to support emerging architectures and cross-architecture parallel-processing-standards, and is working intensely to integrate 
additional architectures, such as ARM low-power processors, into its offline computing and simulation workflows. The collaboration is also engaged in a strong R\&D programme
of artificial intelligence algorithms for reconstruction and data processing and is developing the means to efficiently deploy these algorithms 
at scale and in real-time across its software stack. 

Lastly, LHCb aims at gaining a foothold in precise timing distribution and control. As a consequence of the introduction of timing detectors to fulfil the LHCb Upgrade~II programme, it is essential to attain the capability of operating with $<10\ps$ absolute clock-phase determinism over tens of thousands of channels, hundreds of FPGA-based cards, and multiple years. This will be achieved by utilizing commercial FPGAs combined with dedicated hardware as well as specific techniques in firmware and software for precise and continuous control of the clock at each transmission and reception point.

These pioneering developments in data acquisition, data processing, and time distribution   will ensure that LHCb remains at the forefront of technological developments in HEP over the next decade. Their success will not only help LHCb to maximise the physics reach of Upgrade~II but also chart a path towards a next generation of facilities and experiments, firmly grounded in real-life applications, for the field as a whole. 

\section*{Summary}
The Upgrade~II of LHCb requires considerable technological developments, that will extend in scope and time significantly beyond the end of the developments for the Phase~2 ATLAS and CMS upgrades. 
They will present a stimulating challenge for everyone interested in technologies for future HEP detectors, working in synergy with studies aimed at beyond-LHC applications (Detector Research and Development), while providing a testbed on a live experiment that is expected to be the last large upgrade of the LHC era. 

\addcontentsline{toc}{section}{References}
\bibliographystyle{LHCb}
\bibliography{main,standard,LHCb-PAPER,LHCb-CONF,LHCb-DP,LHCb-TDR, VELO/VELO, UT/UT, MightyTracker/MightyTracker, MagSta/MagSta, RICH/RICH, TORCH/TORCH, PicoCal/PicoCal, Muon/Muon, RTA/RTA, Online/Online, Infrastructure/Infrastructure}
 
\end{document}